# Implementing the PRIDE technique for the JUICE mission at the University of Tasmania

O. J. White[1], G. Molera Calvés[1], and J. Edwards[1]

[1] *School of Natural Sciences, University of Tasmania, Hobart, 7000, TAS, Australia*

**Summary:** The Jupiter Icy Moons Explorer (JUICE) spacecraft is a European Space Agency (ESA) mission to study the Jovian system, currently undergoing an eight-year cruise phase. The Planetary Radio Interferometric and Doppler Experiment (PRIDE) is one of eleven experiments contributing to the JUICE mission. PRIDE aims to conduct radio science experiments using ground-based radio telescopes, through both single dish and Very Long Baseline Interferometry (VLBI) observations. The University of Tasmania (UTAS) is an important contributor to PRIDE and JUICE, though its continental-wide network of radio telescopes in Australia. Over 35 PRIDE observations of JUICE were conducted during the period 2023-24, which enabled detailed analysis of space weather activity in this period, as well as the Lunar-Earth flyby campaign in August 2024. In this paper, we describe PRIDE VLBI observations of JUICE, and the first successful results of PRIDE with the University of Tasmania infrastructure.

## Introduction

Interplanetary spacecraft missions rely on ground-based radio telescopes for command and control operations, and nominal telemetry. Spacecraft radio communications can be used to conduct radio science experiments by utilizing radio telescopes as receivers. These experiments involve analysing various properties of the transmitted signal, including Doppler shift, interplanetary phase scintillation [1], and performing interferometry. PRIDE is a technique developed to enhance the scientific return of spacecraft missions by performing radio science experiments and obtaining precise orbit determination.

The PRIDE technique relies on observing the Doppler shifts of the signal to obtain the spacecraft's radial velocity. Additionally, near-field phase-referencing VLBI is used to measure the spacecraft's lateral position. These measurements can be used to refine the spacecraft state vectors along with a variety of radio science applications including:

- Measuring the interplanetary scintillation between the spacecraft and the Earth, and subsequent determination of the contribution of the ionised plasma from the solar wind and coronal mass ejections [1].
- Modelling of the vertical pressure and temperature profiles of gaseous atmospheres of planets via radio occultation experiments [2].
- Precise measurements of the gravitational field parameters of planetary bodies as well as improving their ephemerides [3].
- Measuring the wind profiles of planetary atmospheres [4].

PRIDE is one of the eleven experiments contributing to the JUICE spacecraft mission [5]. Launched on April 14th, 2023, JUICE commenced an eight-year cruise phase with a series of flybys of the Earth, Moon and Venus. PRIDE-JUICE will support JUICE throughout the cruise phase through precise spacecraft tracking and performing radio science experiments to study space weather, the dynamics of the spacecraft and ensuring the overall health of the mission before reaching the Jovian system. Upon JUICE's arrival in 2031, there will be many milestones in which PRIDE will make measurements of Jupiter and its icy moons. For example, precise spacecraft VLBI tracking will enable analysis of the moons' gravitational field parameters, facilitating investigation of their interiors and their tidal evolution. This is achieved with minimal requirements on the spacecraft's payload, with measurements performed only using the radio signal from the regular spacecraft transmissions [5].

The University of Tasmania (UTAS) is an important member of PRIDE-JUICE through its continental-wide network of radio telescopes. The UTAS Hb (Hobart, Tasmania), Ke (Katherine, Northern Territory) and Yg (Yarragadee, Western Australia) 12m radio telescopes, as well as the Cd (Ceduna, South Australia) 30m, and Ho (Hobart, Tasmania) 26m have been used for these observations. PRIDE Doppler observations have been conducted since its launch in 2023, and PRIDE VLBI phase-referencing observations since 2024.

This paper is divided into three content sections. The Observation Technique section details the principles behind the PRIDE VLBI technique. Next, the PRIDE VLBI implementation at the University of Tasmania is discussed, with the observation methodology and data processing pipeline implemented. Finally, the results of four PRIDE VLBI experiments in July and August 2024 are detailed, including the successful imaging of the spacecraft and natural radio sources, along with the PRIDE Doppler observations.

## Observation Technique

Radio interferometric techniques applied to spacecraft tracking yield estimates of the relative velocity and positions of the spacecraft with respect of the ground stations. If the position of the ground stations is known accurately then these values can be derived for the spacecraft. This section draws from the explanations given in previous radio interferometric studies [5, 6, 7, 8].

Three main data types can be determined from radio observations of spacecraft telemetry: spacecraft range, Doppler shift and interferometric measurements. In the typical mode of operation for VLBI observations, ground stations transmit an uplink signal to a spacecraft, which lock and track it using a phase-locked loop (PLL). This is then phase modulated onto a downlink signal with a fixed frequency ratio. For example, in the C/X-band spacecraft commonly have an uplink frequency of 7145-7190 MHz and a downlink of 8400-8450 MHz [9].

The Doppler shift is the change in frequency between the transmitted and received signal due to the relative motion of the spacecraft (after accounting for the frequency offset of the uplink and downlink signals). This is given by Eqn 1.

$$\Delta f = (\dot{r}/c) f_t \tag{1}$$

where $\dot{r}$ is the spacecraft's instantaneous range rate. This directly yields the spacecraft's radial velocity from the line of sight of the ground station.

There are three different ways in which the signal can be transmitted and received; one-way, two-way and three-way mode, the latter of which is used for PRIDE observations. Three-way mode operates by having two ground stations used, one for transmission and another for reception of the spacecraft signal. Therefore, independent frequency standards are used at the two ground stations, typically hydrogen masers. The stability of the spacecraft Ultra Stable Oscillator (USO) is of the order $10^{-13}$ [10], while the hydrogen masers at the ground station are less than $10^{-14}$ [11], both over $\tau$ = 10s. Since hydrogen masers are an order of magnitude more stable, three-way mode is preferred over one-way for ranging and Doppler observations with VLBI receivers.

The Doppler and range analysis of spacecraft does not give a direct measurement of the spacecraft's angular position. Therefore, errors in the estimated angular position can arise if mistakes are made in the modelling of the forces on the spacecraft affecting the amplitude or phase of the received signal [12]. VLBI however provides direct measurements of the spacecraft's angular position.

The PRIDE technique is a VLBI phase-referencing technique. Connecting radio telescopes in an interferometric array improves the resolution to overcome the limitations of diffraction at the long radio wavelengths. For example, considering Rayleigh diffraction, at an X-band wavelength of 3.6cm, the Hobart 26m radio telescope would have a limiting resolution of approximately 5.8 arcminutes. Meanwhile VLBI baselines of up to 10,000km can enable angular resolutions of up to 1 milliarcsecond.

VLBI observations are conducted through recording at each station separately, then correlating the data post-observation to form an interference fringe pattern. The signals first are shifted to compensate for the signal delay between the stations, which includes the geometric delay $\tau_g$, as well as Doppler, propagation and instrumentation effects. The interferometer setup and fringe pattern formed for a generic two antenna setup is shown in Fig. 1 (left panel).

The output of the correlator can be mathematically described by Eqn 2.

$$r = A_0 |V| \Delta \nu \cos(2\pi \boldsymbol{b}_\lambda \cdot \boldsymbol{k_0} - \Phi_V), \qquad (2)$$

where $A_0$ is the effective antenna area, $V$ is the complex visibility (the Fourier transform of the source brightness), $\Delta \nu$ is the bandwidth, $\boldsymbol{b}_\lambda$ is the baseline vector measured in units of wavelength, $\boldsymbol{k_0}$ is the direction to the image centre and $\Phi_V$ is the error of the visibility phase.

For spacecraft tracking using the VLBI technique, the error in the predicted position of the spacecraft is calculated using the error of the visibility phase $\Phi_V$ [6]. The model used in the correlator needs to account for corrections in the geometric delay $\tau_g$, atmospheric and ionospheric effects, as well as errors in the instrumentation and clocks.

Solving for the phase delay in the interferometer response enables the error in the visibility phase to be determined. In VLBI phase-referencing techniques, such as PRIDE, the phase delay is calculated by observing a phase reference source. Conventionally, fringe-fitting and self-calibration is performed on the target data in VLBI observations, however due to spacecraft having a relatively low SNR and the lack of accuracy around their position and structure, this is not reliable. Instead, the antennas alternate observations between the spacecraft and a quasar, as shown in Fig. 1 (right panel).

Quasars are distant galaxies that appear as bright point-like sources with their positions determined down to an accuracy of several tens of microarcseconds. The quasars are used as reference sources, and the fringe-fitting and self-calibration is performed on them. Thus, if the reference source is within a few degrees of the spacecraft, the delay, delay rate and phase errors of the spacecraft signal can be compensated for. Therefore, any residual error is due to the deviation of the spacecraft's position from its previously calculated position. In this way, the spacecraft's position can be determined.

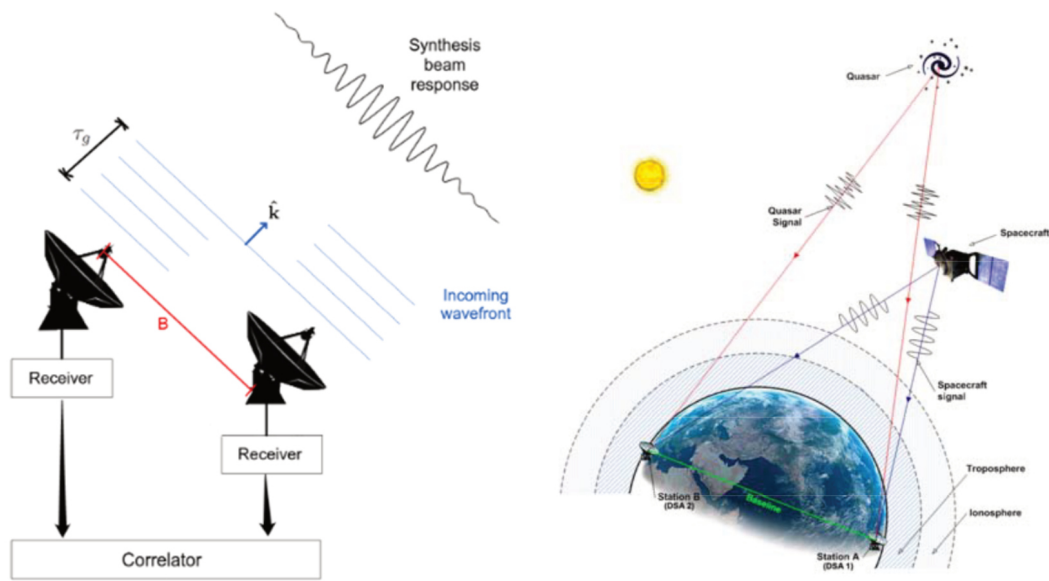

*Figure 1: Left: An interferometer with two antennas, separated by a baseline B and has a time delay between the wavefront reaching the receivers of $\tau_g$. The fringe pattern is shown, with the Full Width Half Maximum of the correlated signal given by 1.22λ/B [6]. Right: Schematic of the VLBI tracking of spacecraft, with a quasar used for phase-referencing [13]*

The PRIDE VLBI phase-referencing observations are scheduled alternating between a quasar calibrator source and the spacecraft, spending up to a few minutes on each. This is unless the quasar is sufficiently close to the spacecraft as to be captured in the primary beam of the antenna, in which case the spacecraft is observed for the entire observation, known as in-beam phase referencing. The geometric delay is then measured by cross correlating the signals between pairs of antennas, which solves for the baseline and the position vectors of the spacecraft. The spacecraft's position is found relative to the nearby quasar source, and ionospheric and atmospheric effects can be compensated for. Additionally, fringe fit scans of a strong quasar source are conducted at the start and end of observations to compensate for the clock offset and rates between stations.

Following observations, in order to measure the relative positions of the spacecraft and the quasar source, the signal from each of these must be correlated separately. For the quasar source, a *far-field delay model* can be utilised to compensate for the geometric delay, as its wavefront can be assumed to be planar when arriving at each station (given they are typically billions of light years distant). For spacecraft communications in the Solar System at X-band the wavefronts have curvature, and a separate *near-field delay model* must be used. These are defined using the Barycentric Celestial Reference System (BCRS), a coordinate system where the origin is defined as the centre of mass of the Solar System.

Following correlation, the output of the broadband correlation of the quasar and spacecraft sources are stored as a FITS Interferometry Data Interchange (FITS-IDI) file. These files can be imported into the Astronomical Image Processing System (AIPS[1]), which allows the parameters necessary for calibration to be calculated. This enables the spacecraft's position to be determined, either by solving the Fundamental Astrometric equation [14] or producing a radio image of the spacecraft, by comparing the offset of its apparent position to the centre of the image, which was the telescope's pointing position.

---

[1] http://www.aips.nrao.edu/aips_faq.html

## PRIDE VLBI Implementation

The observation technique as discussed in the previous section was implemented at the University of Tasmania, using the 12m radio telescopes at Hobart (Hb), Katherine (Ke) and Yarragadee (Yg), as well as the 30m at Ceduna (Cd). The UTAS radio telescopes are equipped with standard VLBI backends, data acquisition systems and hydrogen-masers for the clock standards. These antennas are all capable of receiving in the X-band, which includes JUICE's standard transmission frequency of approximately 8436 MHz.

The 12m antennas' receivers produce signals in linear polarisations (horizontal and vertical), while the transmitted signal is circularly polarised. The radio frequency received by the radio telescopes is down converted, within the 6-10 GHz sky frequency band, to an intermediate frequency between 0-4 GHz. The local oscillator (LO) for the down-conversion is set to 6.00 GHz. The analogue signal is then sampled with the digital baseband converters 3 (dbbc3) [15]. These data use 2-bits per signal, and can have a variable bandwidth, with 8, 16 or 32 MHz typically used for spacecraft observations. The data is stored to disk for offline processing.

The antennas also feature software defined radio (SDR) and GNU Radio[2] to record at a higher number of bits per sample and a lower sampling rate. The design runs in an Ettus X310[3] at Hb and in an Ettus N210 at Ke. These record complex or real data using 32-bits per sample with a 1 MHz bandwidth. Additionally, the Ettus can be used to visualise the spacecraft signal in real-time. Fig. 3 shows the schematics of the current design at the Hobart and Katherine 12m antennas, as well as the Ceduna 30m antenna.

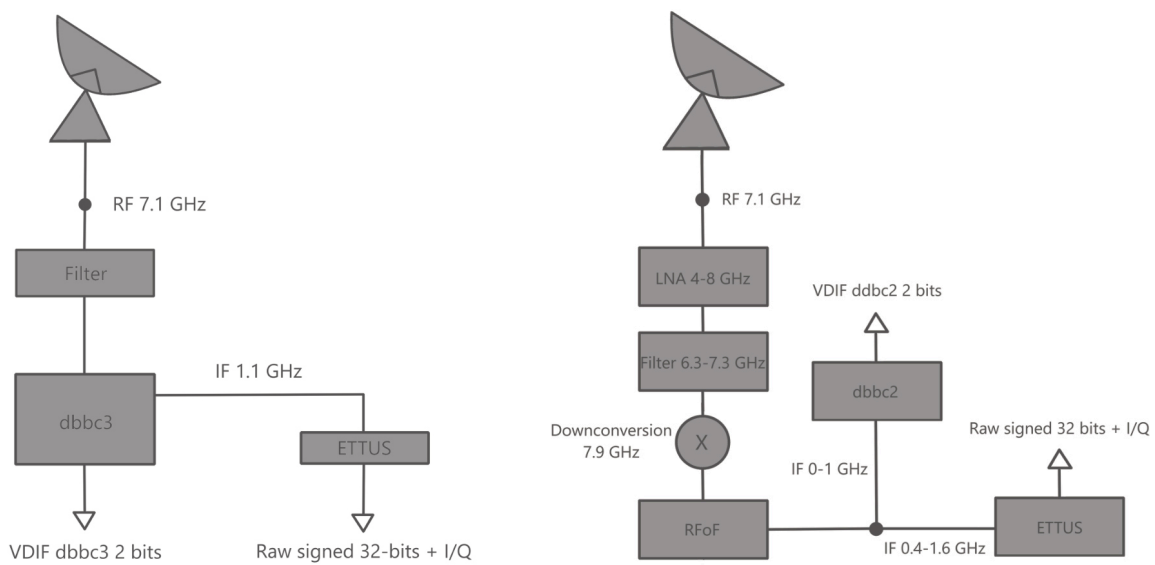

*Figure 3: Diagram of the data recording setup on the Hobart and Katherine 12m antennas (left) and Ceduna 30m (right) [16]. RFoF is Radio Frequency over Fiber.*

Observations are scheduled when JUICE is transmitting, telescope time is available, and the spacecraft is above the horizon at the telescope locations. The JUICE mission operation reports contain a schedule of transmission times. This is used to select an observation time when the spacecraft is communicating with the ESA New Norcia antenna, based in Western Australia, as this provides good elevation overlap with the UTAS antennas.

---

[2] http://www.gnuradio.org/
[3] https://kb.ettus.com/Knowledge_Base

At least two natural sources were needed for each observation: a fringe finder and a phase-reference source. The fringe finder source was ideally a very strong source but could be located anywhere in the visible sky, whilst the phase-reference source was required to be within two degrees of JUICE's position. JUICE's predicted right ascension and declination was found using JPL *Horizons*[4], and the Astrogeo *VLBI Calibrator search*[5] was used to find natural radio sources angularly close to JUICE.

At UTAS, the *Distributed FX* (*DiFX*) software correlator [17] and the *Difxcalc* [18] interferometer model calculator was used to correlate VLBI data post-observation. Previous PRIDE phase referencing experiments used the *SFXC* software correlator [19] and in-house tools developed at the Joint Institute for Very Long Baseline Interferometry European Research Infrastructure Consortium (JIVE) specifically for PRIDE. This is the first attempt at correlating PRIDE experiments with the *DiFX* software correlator [17], the other major software correlator for VLBI used globally.

Correlation for the spacecraft was performed using the near-field version of *DiFX* (version 2.6.2), with the JUICE SPICE *kernel*[6] and NASA's *Planetary Data System Navigation Node*[7] used. The Duev model was used, which solves the light-time equation twice to determine the near-field delay [14]. This was developed as part of the PRIDE package and demonstrated successfully with *SFXC* [3]. The Duev model became integrated into *difxcalc* in 2016, though further verification and comparison has not been conducted. This project began to address this gap. Following correlation in *DiFX*, the data was compiled into a *FITS-IDI* file for further analysis in *AIPS* or *CASA*. It was beyond the scope of this study to implement this stage of the data processing pipeline, however the future implementation at UTAS will use *CASA* [20].

## Results

**Doppler Observations**

Over 35 single antenna experiments of JUICE have been conducted from the University of Tasmania since its launch in April 2023. Observations were scheduled similarly to the PRIDE VLBI, however track the spacecraft continuously. The data for each observation was subsequently processed using the *Spacecraft Doppler Tracker* software (*SDtracker*)[8], along with a series of Python tools (*pysctrack*)[9]. This enables sub-Hz precision detections of the topocentric frequency detections, along with reconstructed and residual phases of the carrier signal [15]. In particular, these observations have applications in the analysis of space weather [1].

The JUICE spacecraft carrier tone along with the residual phases is shown in Figure 4, as well as the topocentric frequency detections versus time in Figure 5.

---

[4] https://ssd.jpl.nasa.gov/horizons/app.html
[5] http://astrogeo.org/calib/search.html
[6] https://s2e2.cosmos.esa.int/bitbucket/projects/SPICE_KERNELS/repos/juice/browse/kernels/spk
[7] https://naif.jpl.nasa.gov/pub/naif/generic_kernels/spk/planets/
[8] https://gitlab.com/gofrito/sctracker
[9] https://gitlab.com/gofrito/pysctrack

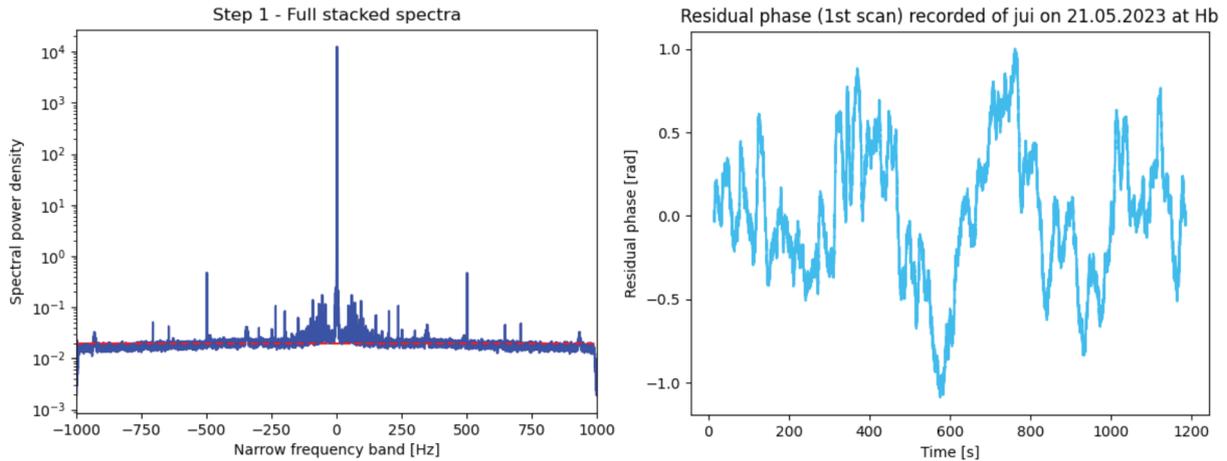

*Figure 4: Left: JUICE spacecraft integrated carrier tone and subtones from the Ceduna 30m radio telescope on 12/10/2023. Right: Residual phases from a 20-minute scan of JUICE from the Hobart 12m antenna on 21/05/2023. At the time of the observation, the spacecraft was at a distance of 0.06 AU and a Sun-Observer-Target elongation angle of 132 degrees.*

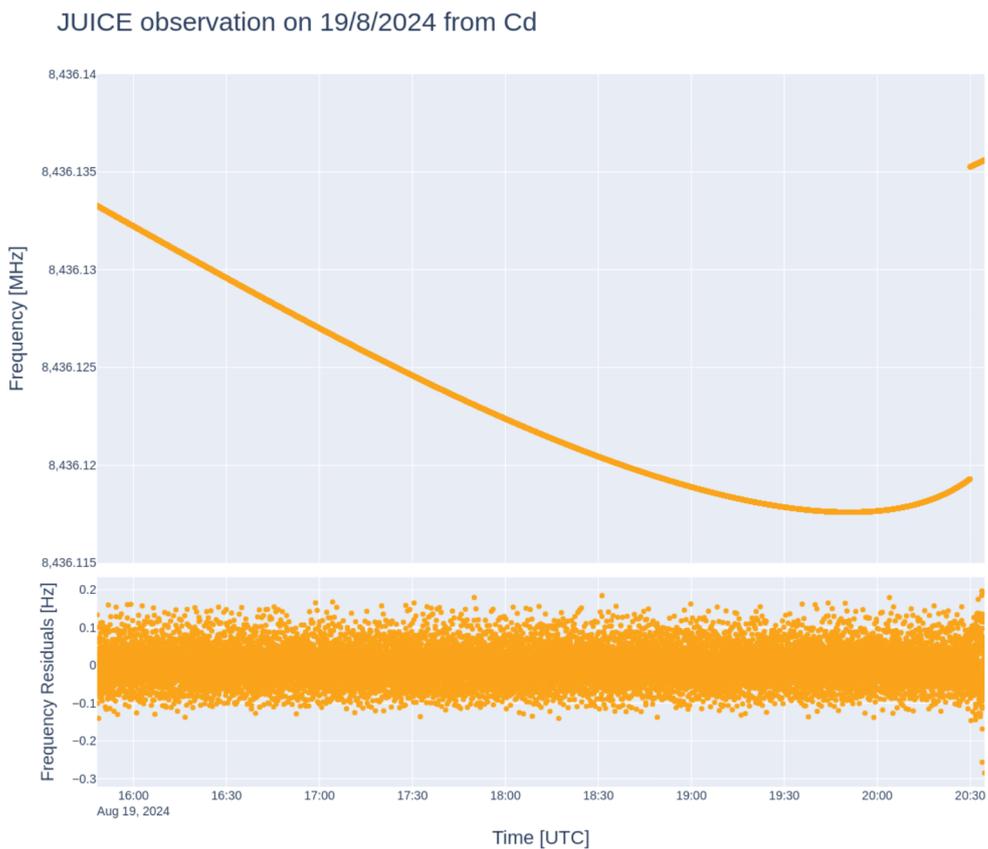

*Figure 5: Revised frequency detections of the JUICE Doppler tracking from the Ceduna 30m antenna on the 19/08/2024, during the Lunar flyby. Transmission switched from two-way to one-way mode just prior to lunar occultation, which resulted in the frequency jump before it was lost soon after 20:30 UTC.*

*Table 1: Summary of PRIDE Doppler observations of JUICE from the University of Tasmania. Observations were conducted in X-band in three-way mode. Distance and SOT values obtained from Horizons.*

| Epoch | Stations | Distance (AU) | Sun-Observer-Target Elongation Angle (deg) |
| --- | --- | --- | --- |
| 14/04/2023 | Ho | 0.0003 | 134.1 |
| 17/04/2023 | Ho, Hb, Ke | 0.0054 | 157.4 |
| 20/04/2023 | Ho, Hb, Ke | 0.0100 | 155.5 |
| 21/04/2023 | Ho | 0.0115 | 154.7 |
| 22/04/2023 | Ho, Hb, Ke, Cd | 0.0131 | 153.9 |
| 24/04/2023 | Ho, Hb, Ke, Yg, Cd | 0.0161 | 152.3 |
| 26/04/2023 | Hb | 0.0192 | 150.6 |
| 02/05/2023 | Ke | 0.0285 | 145.7 |
| 04/05/2023 | Hb, Ke | 0.0316 | 144.1 |
| 08/05/2023 | Ho, Hb, Ke, Yg | 0.0380 | 141.0 |
| 09/05/2023 | Ho, Yg | 0.0396 | 140.2 |
| 10/05/2023 | Hb, Ke | 0.0412 | 139.4 |
| 11/05/2023 | Hb, Ke | 0.0428 | 138.7 |
| 13/05/2023 | Hb, Ke | 0.0461 | 137.2 |
| 14/05/2023 | Hb, Ke | 0.0477 | 136.4 |
| 15/05/2023 | Ho, Hb, Ke | 0.0494 | 135.7 |
| 20/05/2023 | Hb, Ke, Cd | 0.0579 | 132.1 |
| 21/05/2023 | Hb, Cd | 0.0597 | 131.5 |
| 22/05/2023 | Ke, Cd | 0.0614 | 130.8 |
| 27/05/2023 | Hb, Ke | 0.0702 | 127.5 |
| 02/06/2023 | Ho, Hb, Ke | 0.0808 | 123.8 |
| 03/06/2023 | Hb, Ke | 0.0826 | 123.2 |
| 08/06/2023 | Ke | 0.0916 | 120.3 |
| 16/06/2023 | Ho | 0.1060 | 115.9 |
| 19/06/2023 | Hb, Ke | 0.1114 | 114.4 |
| 21/06/2023 | Hb, Ke, Cd | 0.1149 | 113.4 |
| 22/06/2023 | Hb, Ke | 0.1167 | 112.9 |
| 23/06/2023 | Hb | 0.1185 | 112.4 |
| 26/06/2023 | Hb, Ke | 0.1237 | 110.9 |
| 05/07/2023 | Hb | 0.1390 | 106.7 |
| 06/07/2023 | Hb | 0.1406 | 106.2 |
| 12/07/2023 | Hb | 0.1501 | 103.5 |
| 11/09/2023 | Hb, Ke, Cd | 0.1943 | 78.3 |
| 14/09/2023 | Hb | 0.1931 | 76.9 |
| 12/10/2023 | Cd | 0.1654 | 60.7 |
| 12/11/2023 | Ho, Cd | 0.1217 | 23.9 |
| 14/08/2024 | Hb, Ke, Yg | 0.0119 | 165.6 |
| 19/08/2024 | Cd | 0.0029 | 176.0 |
| 20/08/2024 | Hb, Ke, Yg, Cd | 0.0008 | 169.6 |
| 21/08/2024 | Cd | 0.0017 | 80.9 |
| 24/08/2024 | Cd | 0.0058 | 84.5 |

**PRIDE VLBI**

Four PRIDE VLBI observations were conducted of JUICE in the July-August 2024 period preceding the Lunar-Earth flyby, on July 2nd, August 3rd, 13th and 19th. The fringe finder and phase reference sources used for each observation are listed in Table 2.

*Table 2: Radio sources used for PRIDE VLBI observations. The flux densities are approximate, obtained from the Astrogeo Center, given in Janskys.*

| Source | Flux density (Jy) | Type | Sessions |
|---|---|---|---|
| **B1921-293** | >5 | Fringe Finder | July 2nd, August 3rd, 19th. |
| **J2211-1328** | >0.3 | Phase Reference | July 2nd, August 19th. |
| **J2211-1150** | 0.1 | Phase Reference | July 2nd. |
| **B2255-282** | >2 | Fringe Finder | August 3rd, 13th. |
| **J2229-0832** | >1.5 | Phase Reference | August 3rd, 13th. |
| **J2240-0836** | 0.1 | Phase Reference | August 3rd. |

Fringe fitting was performed for the August 3rd and July 2nd observations. Both observations used the Hobart, Katherine and Yarragadee 12m antennas, whilst the July 2nd observation also featured Ceduna. Following this, the phase calibrator scans were correlated. As shown in Table 1, two phase calibrators were used for each experiment. For the August 3rd observation, the J2229-0832 scans were all detected, though the J2240-0836 scans were too weak to be detected along any baseline for the scan lengths used. Similarly for the July 2nd observation, J2211-1150 was too weak to be detected, whilst J2211-1328 was stronger, though only definitively detected along baselines including the larger Cd antenna. The single band delay was nominal on all of the detected scans, indicating a successful fringe fitting.

The August 3rd observation was used for subsequent spacecraft correlation due to the successful natural source correlations along all baselines. The Duev, Sekida & Fukushima, and spacecraft ranging models [21] were all used separately to generate the near-field model, with negligible differences in the single band delay of less than 0.1 nanoseconds, though the Duev model was used for the final product. The correlation had a significant single band delay despite the nominal single band delay in the phase calibrator scans. For example, the first JUICE scan had a single band delay of 251.149 ns along the Hb-Ke baseline with left-left polarisation, whilst the phase calibrator scans had single band delays of less than 0.01 μs.

After testing a numerous combination of spectral resolutions, integration times and kernels with negligible change, the source of error is likely to have arisen either from the error in the telescope pointing offset compared to the spacecraft's actual position, or errors in the SPICE JUICE *kernel* due to inaccuracies in the orbital modelling. Further experiments and validation of the spacecraft positions in *CASA/AIPS* will determine which of these errors has caused this.

*AIPS* was used to produce images of the spacecraft and the phase calibrator source. An image of the phase calibrator source is shown in Fig. 4. The image shows the compactness of the source and its suitability to be used as a reference source for these types of experiments. An image of JUICE is shown in Fig. 5. In the plot, the power spectra of the signal have been overlaid to the dirty image of the spacecraft. The spacecraft is far away from the phase centre, due to errors on the orbit. That is expected, since the number of observations and tracking data is limited during the cruise phase.

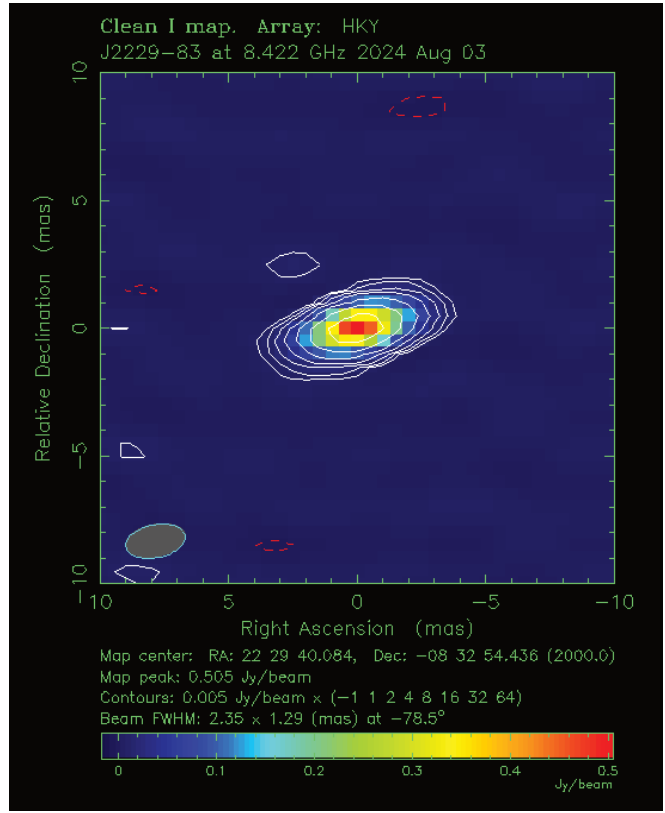

*Figure 4: Image of J2229-0832 produced from the phase calibrator scans of the August 3rd experiment with the Aid of Dr Sándor Frey. Hb, Ke and Yg were used, which shows the resolved source at the expected position with an extended feature and minimal structure.*

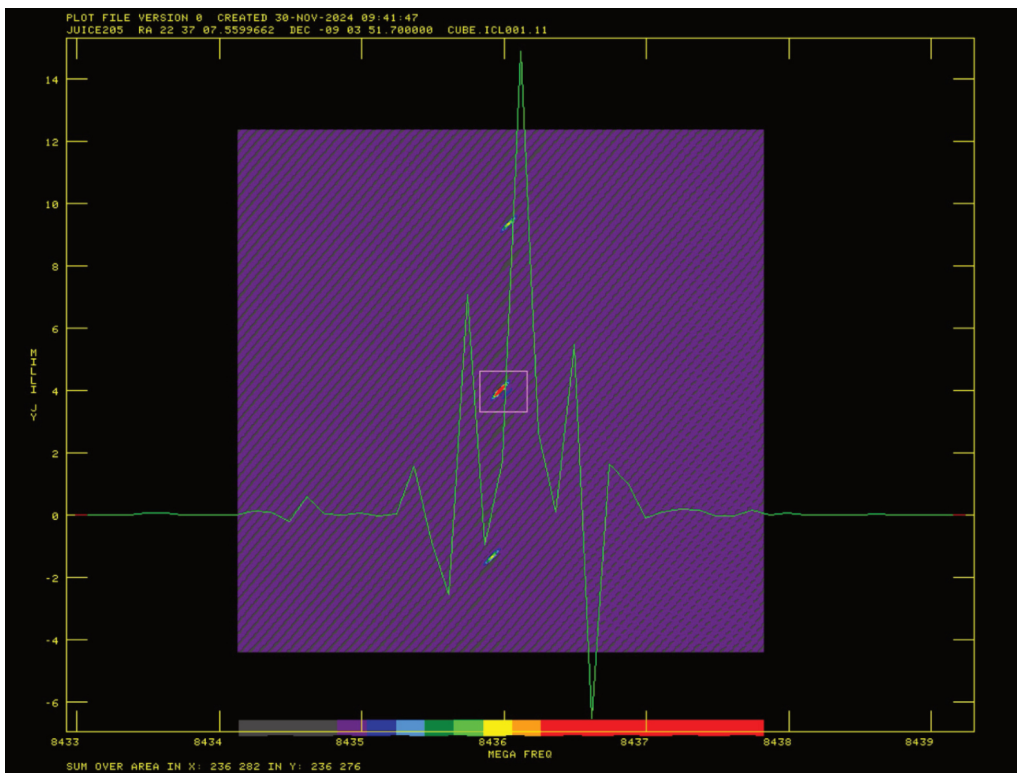

*Figure 5: Image of JUICE produced from the August 3rd observation with the aid of Dr Ross Burns. The spacecraft image is displayed with an overlay of its spectra, with the narrow tone at the expected spacecraft frequency confirming the source as JUICE.*

Previous studies imaging spacecraft with the PRIDE VLBI technique were published only two years after observations were taken, due to the complexity of the observations and the development of the technique [2, 14]. We have achieved the data compilation, processing and production of images in only 3 months, with a plethora of software packages used.

All of the calibrations and corrections from the Observation Technique section have been conducted in order to produce these images. We detected JUICE despite its coordinates being erroneous by several kilometres, which is normal during cruise phase; spacecraft orbits become much more precise when orbiting planets. However, this further demonstrates the ability of PRIDE to contribute to the orbital modelling. Work is ongoing to precisely determine the magnitude of these errors and incorporate it into the orbit predictions.

## Conclusion

We have implemented pipelines for various components of the PRIDE VLBI technique at UTAS. The scheduling and observing techniques using the UTAS antennas have been fully implemented, along with the correlation component of the data processing pipeline. Four experiments were conducted for the JUICE spacecraft during July-August 2024. Additionally, 35+ PRIDE Doppler experiments were conducted in the 2023-2024 period, which will enable analysis of the spacecraft's space weather environment. The last step in producing the desired data output of the spacecraft radio images and determination of the state vectors is in analysing the correlator output in *CASA* or *AIPS*. With the assistance of Dr Sándor Frey and Dr Ross Burns these images were produced using *AIPS*, with the full data processing achieved in only three months despite the complexity of the VLBI technique and the various software packages used. A goal of future research at UTAS will be to process the data using *CASA*.

## Acknowledgements

Thank you to Ahmad Jaradat, David Schunck, Dr Tiege McCarthy, Dr Sándor Frey and Dr Giuseppe Cimo for providing their expertise and advice in processing the PRIDE VLBI data of JUICE using the DiFX software correlator. Dr Sándor Frey and Dr Ross Burns gave their assistance in producing images of the calibrator sources and JUICE respectively using AIPS.